\documentclass[aps,showpacs,prep,twocolumn]{revtex4}
\usepackage{amssymb}
\usepackage{amsmath}
\usepackage{bm}

\begin{document}

\title{$\Lambda(t)$CDM and the present accelerating expansion of the universe from 5D scalar vacuum}
\author{ $^{1}$ Jos\'e Edgar Madriz Aguilar, $^{2}$ J. Zamarripa, $^{3}$ A. Peraza and $^{1}$ J. A. Licea\thanks{%
E-mail address: edgar.madriz@red.cucei.udg.mx} }
\affiliation{ $^{1}$ Departamento de Matem\'aticas, \\
Centro Universitario de Ciencias Exactas e Ingenier\'{i}as (CUCEI),\\
Universidad de Guadalajara. \\
Av. Revoluci\'on 1500 S.R. 44430,\\
Guadalajara, Jalisco, M\'exico.\\
e-mail: madriz@mdp.edu.ar,  jlicea@itesm.mx\\
\\
$^{2}$ Centro Universitario de los Valles,\\
Universidad de Guadalajara,\\
Carretera Guadalajara-Ameca km 45.5, CP 46600, Ameca, Jalisco, M\'exico.\\
e-mail: zama\_92@live.com.mx 
\\
and \\
$^{3}$ Departamento de F\'{i}sica,\\
Centro Universitario de Ciencias Exactas e Ingenier\'{i}as (CUCEI),\\
Universidad de Guadalajara. \\
Av. Revoluci\'on 1500 S.R. 44430,\\
Guadalajara, Jalisco, M\'exico.\\
e-mail: aperaza6@hotmail.com }

\begin{abstract}
	
In this letter we investigate some consequences of considering our 4D observable universe as locally and isometrically embeded into a 5D spacetime, where gravity is described by a Brans-Dicke theory in vacuum. Once we impose the embeding conditions we obtain that gravity on the 4D spacetime is governed by the Einstein field equations modified by an extra term that can play the role of a dynamical cosmological constant. Two examples were studied. In the first we derive a cosmological model of a universe filled only with a cosmological constant. In the second we obtain a cosmological solution describing a universe filled with matter, radiation and a dynamical cosmological constant, in agree with observational data. Due to the 5D geometrical origin for the dynamical cosmological constant we interpret that the acceleration in the expansion is explained without the introduction of a dark energy component. Moreover, all 4D matter sources are geometrically induced in the same manner as it is usually done in the Wesson's induced matter theory. \\

\end{abstract}

\maketitle

\vskip .5cm


\section{Introduction}

The present accelerating expansion of the universe has been a topic of modern cosmology since its discovery in 1998 \cite{Rebi1,Rebi2,Rebi3}. The simplest theoretical explanation is given by the concordance model $\Lambda$CDM, where dark matter is regarded as a cold pressureless fluid whereas the dark energy, which is responsible for such acceleration, is driven by a cosmological constant $\Lambda$. The success of this model is due to its fitting to observational data \cite{Rebi4,Rebi5}. However, it is not free of problems. The main problem of this model relies in the origin and the smallness of the cosmological constant, usually called the cosmological constant problem \cite{Rebi5}. Basically, the $\Lambda$CDM model fails in explaining the huge difference between the cosmological constant value needed to explain the present accelerating expansion in concordance with observational data ($\Lambda\simeq 10^{-52}\, m^{-2}$), and the value obtained from quantum field theory ($\Lambda\simeq 10^{122}\, m^{-2}$), when it is associated with the vacuum fluctuations of quantum fields \cite{Rebi6}.\\

The proposals for solving the cosmological constant problem are based in general on two lines of reasoning. In the first one it is postulated the existence of an exotic component in the material content of the universe i.e. a dark energy energy component. Models with scalar fields and $\Lambda (t)$CDM models, among others, have been proposals within this line of reasoning \cite{Rebi7,Rebi8,Rebi9}. In the other one the acceleration of the expansion of the universe has its origin in gravity itself. In this line enter for example modified theories of gravity and theories of gravitation in extra dimensions \cite{Rebi10,Rebi11,Rebi12}.\\

Physical theories in more than four dimensions emerged with the idea of unification of fundamental interactions. However, with the pass of time, the motivations range has increased. In 1914 Gunnar Nordstrom starts these ideas with the introduction of one extra coordinate in his scalar gravity theory, which attempt to unify gravity with electromagnetism \cite{Rebi13}. Subsequently, with the same motivation, Kaluza and Klein introduced a compact extra dimension in their model. The Arkani-Hamed-Dimopoulos-Dvali brane models use large extra dimensions to explain the weakness of gravity compared with the rest of fundamental interactions, issue known as the hierarchy problem \cite{Rebi14,Rebi15}. The Randall-Sundrum brane world models appeared also motivated by the hierarchy problem. In these models they propose a solution to the hierarchy problem from a 5D anti-De-Sitter spacetime with the extra dimension considered compact. Since the emergence of these brane world models many others have appeared increasing the famous zoo of brane world models, even in more than five dimensions. In brane worlds our universe is represented by a 4D hypersurface, called the brane, embedded in a higher dimensional spacetime called the bulk \cite{Rebi16,Rebi17}.\\

Currently, an interesting feature shared among brane world models is that the electroweak and strong interactions of the standard model of particle physics, are confined to the brane while gravity can propagate through the bulk. Hence, gravitons propagate along the extra coordinate and the gravitational coupling in the bulk can depend on the extra coordinate, even when on the brane it becomes a constant ($G=M_{p}^{-2}$) \cite{Rebi14,Rebi15}. It follows from this fact, that a higher dimensional Brans-Dicke theory can be a natural framework to describe the propagation of gravitons along the bulk, because gravitational coupling in Brans-Dicke theory is depending on the spacetime coordinates \cite{Rebi18,Rebi19,Rebi20}. \\

The fact that gravity can propagate along the higher dimensional space is not exclusive of brane world scenarios. In general, theories known as non-compact Kaluza-Klein theories, like for instead the induced matter theory, also share this property. This theory was introduced by Paul Wesson and collaborators \cite{Rebi12}. The main idea in this theory is that the 5D bulk is assumed Ricci flat  ($\,^{(5)}\! R_{ab}=0$), and solutions of these field equations can describe on our 4D universe the Einstein field equations of general relativity, where the soruces of matter are determined (or induced) by the 5D geometry. Tecnically, the extra dimension is considered non-compact and it is responsible for the presence of matter sources in our 4D universe. A plenty of contributions have been made in this approach, as for example inflationary cosmology \cite{Rebi21, Rebi22}, gravitational waves \cite{Rebi23}, induced matter with Brans-Dicke theory \cite{Rebi24} and $f(R)$ theories \cite{Rebi25}, among many others.\\

In this letter we obtain cosmological solutions of a 5D Brans-Dicke theory of gravity in vacuum, that describe our 4D universe as embeded into the 5D spacetime, where the accelerating cosmic expansion is explained without the need of the introduction of a dark energy component, instead the acceleration is associated with the fact that the gravitational coupling depends only of the extra coordinate along the 5D Bulk.\\

The letter is organized as follows. We start by giving a little introduction in section I. In the section II we develop the general formalism. In section III we show as an example a cosmological scenario with a universe filled only with a cosmological constant. Section IV is devoted to a cosmological setting with matter and radiation in which the acelerated expansion of the universe is generated by a dynamical cosmological constant of geometrical origin. We leave section V for final comments.\\

Our conventions are Latin indices run in the range $(0,1,...,4)$ with the exception of $i$ and $j$ that take values in the range $(1,2,3)$. Greek indices run from $(0,1,2,3)$. The metric signature we use is $(+,-,-,-,-)$. Finally we adopt units on which the speed of light $c=1$.

\section{The Formalism}

Let us start considering a five-dimensional (5D) spacetime in absence of matter sources, where gravity is described by a Brans-Dicke (BD) theory. The action for a 5D BD theory of gravity in vacuum,  without a potential term, is in this case \cite{Rebi24}
\begin{equation}\label{a1}
^{(5)}{\cal S}=\int d^{5}\xi\sqrt{g_5}\left[\varphi\, ^{(5)}\!R-\frac{\omega}{\varphi}g^{ab}\varphi_{,a}\varphi_{,b}\right], 
\end{equation}
where $\varphi(\xi)$ is the BD scalar field which is related with the 5D dynamical gravitational coupling, $\omega$ is the BD constant parameter and $g_5$ is the determinant of the 5D metric $g_{ab}$.  On a coordinate chart $\lbrace x^{\mu},\psi\rbrace$, the differential line element can be splitted in the form
\begin{equation}\label{a2}
ds_5^{2}=g_{\mu\nu}(x,\psi)dx^{\mu}dx^{\nu}+g_{\psi\psi}(x,\psi)d\psi^{2},
\end{equation}
being $\psi$ the space-like non-compact extra dimension. The Einstein-Hilbert variational procedure to the action (1) generates the field equations 
\begin{small} 
\begin{eqnarray}\label{a3}
&&^{(5)}\!G_{ab}= \frac{\omega}{\varphi ^2}\left[\varphi _{,a}\varphi_{,b}-\frac{1}{2}g_{ab}\varphi^{,c}\varphi_{,c}\right]+\frac{1}{\varphi}\left[\varphi_{;a;b}-g_{ab}\varphi^{;c}\,_{;c}\right],\nonumber \\ \\
\label{a4}
&&^{(5)}\!\Box\varphi =0,
\end{eqnarray}
\end{small}
where the semicolon is denoting 5D covariant derivative and $^{(5)}\Box$  denotes the 5D d'Alembertian operator. Inserting (\ref{a4}) in (\ref{a3}) we obtain
\begin{equation}\label{ab1}
^{(5)}G_{ab}= \frac{\omega}{\varphi ^2}\left[\varphi _{,a}\varphi_{,b}-\frac{1}{2}g_{ab}\varphi^{,c}\varphi_{,c}\right]+\frac{\varphi_{;a;b}}{\varphi}.
\end{equation}
In our approach, the universe we live in is represented by a generic 4D hypersurface $\Sigma:\psi=\psi_0$, obtained after taking a folitation of the whole 5D space by choosing $\psi=\psi_0$. The induced metric on $\Sigma$ can be written as
\begin{equation}\label{a5}
ds^{2}=h_{\alpha\beta}(x)dx^{\alpha}dx^{\beta},
\end{equation}
where $h_{\alpha\beta}=g_{\alpha\beta}(x,\psi_0)$. The central idea of this letter is to investigate the possibility to obtain solutions of the 5D BD field equations (\ref{a3}), that allow to recover on $\Sigma$ the field equations of general relativity characterized by a constant gravitational coupling, without the necessity to use the well-known passage between the Jordan and Einstein frames \cite{Rebi20} i.e. we will use only a dimensional reduction mechanism. Thus, as the gravitational coupling in 4D must be a constant, it follows that the solutions we are looking for, must obey in 5D the condition: $(16\pi G_5)^{-1}=\varphi(\psi)$ while on $\Sigma$ they must satisfy: $(16\pi G)^{-1}=\varphi(\psi_0)$.\\

If we assume separability of the metric functions $g_{\mu\nu}(x,\psi)$ and $g_{\psi\psi}(x,\psi)$ in the form
\begin{eqnarray}\label{abc1}
g_{\mu\nu}(x,\psi)&=&h_{\mu\nu}(x){\cal A}^{2}(\psi),\\
\label{abc2}
g_{\psi\psi}(x,\psi)&=&\epsilon \tilde{g}_{\psi\psi}(x,\psi)=\epsilon\Phi^2(x){\cal P}^2(\psi), 
\end{eqnarray}
with $\epsilon=\pm 1$ depending of the metric signature. The equation (\ref{a4}) gives for solution
\begin{equation}\label{abc3}
\varphi(\psi)=C_1\int\frac{{\cal P}(\psi)}{{\cal A}^4(\psi)}d\psi + C_2,
\end{equation}
where $C_1$ and $C_2$ are integration constants. It is easy to see from (\ref{abc3}) that depending of the form of ${\cal P}(\psi)$ and ${\cal A}(\psi)$, the desired condition $\varphi(\psi_0)=(16\pi G)^{-1}$ can be in principle satisfied, thus establishing on certain manner the existence of the solutions we are interested in.\\

Now, by using the Gauss-Codazzi-Ricci equations (see for example \cite{Rebi12} ) the dynamical field equations (\ref{ab1}) induced on $\Sigma$ read
\begin{eqnarray}
^{(4)}G_{\alpha\beta}&=&\frac{1}{2\varphi(\psi_0)}\,T_{\alpha\beta}^{(IM)}+\frac{1}{2\varphi(\psi_0)}E_{\alpha\beta}^{(\varphi)}\nonumber\\
\label{a11}
&-&\frac{1}{2\varphi(\psi_0)}\left[h^{\mu\nu}E_{\mu\nu}^{(\varphi)}-\frac{2}{3}E^{(\varphi)}\right]h_{\alpha\beta},
\end{eqnarray} 
where the auxiliar tensor $E_{ab}^{(\varphi)}$ is defined by
\begin{eqnarray}\label{abc4}
E_{ab}^{(\varphi)}&=&\frac{\omega}{\varphi^2}\left[\varphi_{,a}\varphi_{,b}-\frac{1}{2}g_{ab}\varphi^{,c}\varphi_{,c}\right]+\frac{\varphi_{;a;b}}{\varphi},\\
\label{abc5}
E^{(\varphi)}&=&g^{ab}E_{ab}^{(\varphi)},
\end{eqnarray}
where $T^{(\varphi)}=g^{ab}T_{ab}^{(\varphi)}$ and $T^{(IM)}_{\alpha\beta}$ is the usual energy momentum tensor for induced matter which is given by the expression \cite{Rebi12}
\begin{small}
\begin{eqnarray}
&&T^{(IM)}_{\alpha\beta}=g^{\psi\psi}(g_{\psi\psi ,\alpha})_{;\beta}-\frac{1}{2}(g^{\psi\psi})^{2}\left[g^{\psi\psi}\overset{\star}{g}_{\psi\psi}\overset{\star}{g}_{\alpha\beta}-\overset{\star\star}{g}_{\alpha\beta}\right.\nonumber\\
&& \left.+g^{\lambda\mu}\overset{\star}{g}_{\alpha\lambda}\overset{\star}{g}_{\beta\mu} -\frac{1}{2}\,g^{\mu\nu}\overset{\star}{g}_{\mu\nu}\overset{\star}{g}_{\alpha\beta}+\frac{1}{4}\,g_{\alpha\beta}\left(\overset{\star}{g}^{\mu\nu}\overset{\star}{g}_{\mu\nu}+(g^{\mu\nu}\overset{\star}{g}_{\mu\nu})^{2}\right)\right].\nonumber\\
\label{a12} 
\end{eqnarray}
\end{small}
For the particular case where the BD scalar field $\varphi$ depends only on the extra coordinate, the effective 4D field equations (\ref{a11}) reduce to
\begin{equation}\label{a13}
^{(4)}G_{\alpha\beta}=\frac{1}{2\varphi(\psi_0)}T_{\alpha\beta}^{(IM)}+\Lambda(x)h_{\alpha\beta}
\end{equation}
where $\Lambda(x)\equiv [\omega/(2\varphi(\psi_0)^{2})](g^{\psi\psi}\overset{\star}{\varphi}^{2})|_{\psi=\psi_0}$. When the condition $\varphi(\psi_0)=(16\pi G)^{-1}$ is satisfied, the effective 4D field equations (\ref{a13}) are the field equations of general relativity with the extra term $\Lambda(x)h_{\alpha\beta}$. Matter configurations can be classically described by the energy momentum tensor $T_{\alpha\beta}^{(IM)}$ as it is usually done in the induced matter approach  \cite{Rebi12}. The auxiliar function $\Lambda(x)$ can be interpreted as a correction to the field equations of general relativity coming from the 5D BD scalar field and of the 5D geometry through $g_{\psi\psi}(x,\psi)$. Thus if $g_{\psi\psi}$ is a constant of the spacetime coordinates $\lbrace x^\alpha \rbrace$, then $\Lambda(x)$ will become a constant $\Lambda_{0}=(\omega/(2\varphi(\psi_0)^2))g_{0}^{\psi\psi}\overset{\star}{\varphi}^2(\psi_0)$ and if $g_{\psi\psi}$ has time dependence then $\Lambda(x)$ will be $\Lambda(t)=\omega/(2\varphi(\psi_0)^2)g^{\psi\psi}(t)\overset{\star}{\varphi}^2(\psi_0)$.
These two options open the possibility to explain the present accelerating expansion of the universe without the introduction of dark energy. Instead we explain  the present expansion of the universe as a consequence of two facts: gravity propagates through the fifth extra dimension and the 5D gravitational coupling is dynamical along the extra coordinate.\\

Moreover, the $\Lambda$CDM cosmological model can be, in principle, derived as a solution of the 5D field equations when $g^{\psi\psi}$ is a constant, whereas some $\Lambda(t)$CDM models can also be obtained when $g^{\psi\psi}$ is an acceptable funtion of time. Something very interesting of our model is that in here, the ``cosmological constant´´$\Lambda_0$ has no the problems of the standard cosmological constant due to the fact that it is not associated with the vacuum energy of quantum fields, instead it is consequence of the 5D bulk geometry.

\section{The  cosmological constant induced from 5D warped product spaces}

In order to illustrate how the geometrical mechanism works to induce a cosmological constant on the 4D spacetime $\Sigma _0$ we consider the 5D line element 
\begin{equation}\label{b1}
dS^{2}=e^{2A(\psi)}[dt^{2}-a^{2}(t)\delta_{ij}dx^{i}dx^{j}]-d\psi^{2},
\end{equation}
where $t$ is the cosmic time, $A(\psi)$ is a warping factor, $a(t)$ is the cosmological scale factor and $\delta_{ij}$ is the Kronecker delta. On this geometrical background the field equations (\ref{ab1})  read
\begin{small}
\begin{eqnarray}
\label{b2}
&&3H^{2}-3(\overset{\star}{A}^{2}+\overset{\star\star}{A})e^{2A}= \frac{1}{2}\frac{\overset{\star}{\varphi}}{\varphi}\left(\omega\frac{\overset{\star}{\varphi}}{\varphi}-2\overset{\star}{A}\right)e^{2A},\\
\label{b3}
&&2\frac{\ddot{a}}{a}+H^{2}-3(\overset{\star}{A}^{2}+\overset{\star\star}{A})e^{2A}=\frac{1}{2}\frac{\overset{\star}{\varphi}}{\varphi}\left(\omega\frac{\overset{\star}{\varphi}}{\varphi}-2\overset{\star}{A}\right)e^{2A},\\
\label{b4}
&&-3\frac{\ddot{a}}{a}-3H^{2}+6\overset{\star}{A}^{2}e^{2A}=\frac{1}{2}\left[\omega\left(\frac{\overset{\star}{\varphi}}{\varphi}\right)^{2}+\frac{\overset{\star\star}{\varphi}}{\varphi}\right]e^{2A}.
\end{eqnarray}
\end{small}
The equations (\ref{b2}), (\ref{b3}), (\ref{b4}) correspond respectively to the components $^{(5)}G_{tt}$, $^{(5)}G_{ii}$ and $^{(5)}G_{\psi\psi}$. Thus the combination $^{(5)}G_{tt}+3\,^{(5)}G_{ii}+2\,^{(5)}G_{\psi\psi}$ yields 
\begin{equation}\label{b5}
2\varphi\overset{\star\star}{\varphi}-4\overset{\star}{A}\varphi\overset{\star}{\varphi}+3\omega\overset{\star}{\varphi}^{2}+12\overset{\star\star}{A}\varphi ^{2}=0.
\end{equation}
The equation (\ref{abc3}) under the geometrical background (\ref{b1}) gives 
\begin{equation}\label{b6}
\varphi(\psi)=k_{\psi}\int_{\psi_0}^{\psi}\exp{[-4A(\psi\prime)]}\,d\psi\prime +\varphi_{0},
\end{equation}
where $\varphi _{0}=\varphi(\psi_0)$ and $k_{\psi}$ is a separation constant. Both equations (\ref{b5}) and (\ref{b6}) determine the Brans-Dicke scalar field $\varphi$ as a function of the warping factor $A(\psi)$. Hence for a warping of the form 
\begin{equation}\label{b7}
A(\psi)=n\,\ln\left[\frac{\psi_0}{\psi+\psi_0}\right],
\end{equation}
being $n>0$, the expression (\ref{b6}) leads to a Brans-Dicke scalar field
\begin{equation}\label{b8}
\varphi(\psi)=\frac{1}{4n+1}\frac{k_\psi}{\psi_{0}^{4n}}\,(\psi+\psi_0)^{1+4n}\,.
\end{equation}
Inserting the equation (\ref{b8}) in (\ref{b5}) it gives that in order to (\ref{b5}) to be satisfied for the field (\ref{b8}) necessarily 
\begin{equation}\label{b9}
\omega=-\frac{8n(2n+1)}{(4n+1)^{2}}.
\end{equation}  
On the other hand in order to the condition $\varphi(\psi _0)=1/(16\pi G_N)$ holds, the separation constant in (\ref{b8}) must be given by $k_\psi=[(4n+1)/(16\pi G_N)][1/(\psi_{0}2^{4n+1})]$ and thus the expression (\ref{b8}) finally becomes
\begin{equation}\label{b10}
\varphi(\psi)=\frac{1}{16\pi G_{N}}\left(\frac{\psi +\psi_0}{2\psi_{0}}\right)^{1+4n}.
\end{equation}
The effective 5D gravitational coupling is given as in the usual Brans-Dicke theory of gravity by $^{(5)}G_{eff}=1/\varphi$ and hence according to  (\ref{b10}) the 5D effective gravitational coupling reads
\begin{equation}\label{b11}
G_{eff}(\psi)\equiv\frac{1}{\varphi(\psi)}=16\pi G_{N}\left(\frac{2\psi_0}{\psi+\psi_0}\right)^{1+4n}\,.
\end{equation}
The induced metric (\ref{a5}) on the generic 4D hypersurface $\Sigma _0$ derived from (\ref{b1}) can be written as 
\begin{equation}\label{be1}
ds^{2}=dT^{2}-a^{2}(T)\delta_{ij}dX^{i}dX^{j},
\end{equation}
where we have made the space-time coordinate rescaling $dT=e^{A_0}dt$, $dX^{i}=e^{A_0}dx^{i}$. Clearly this metric corresponds to a FRW metric with a scale factor $a(t)$. According to (\ref{a13}) and employing the equations (\ref{b9}) and (\ref{b10}) the induced 4D cosmological constant will be in this case
\begin{equation}\label{b12}
\Lambda_{0}=\frac{n(1+2n)}{\psi_{0}^{2}}.
\end{equation}
The induced matter on 4D is described classically by the energy momentum tensor given by (\ref{a12}), however as it is usually done in cosmological applications on the framework of the induced matter approach \cite{Rebi12}, we can regard matter in 4D as a perfect fluid through an energy momentum tensor of the form $T_{\mu\nu}=(\rho_{(IM)}+p_{(IM)})u_{\mu}u_{\nu}-p_{(IM)}h_{\mu\nu}$ with the 4-velocity of the observers given by $u^{\mu}=\delta^{\mu}_{0}$ (comoving observers located on $\Sigma _0$) and satisfying $u^{\mu}u_{\mu}=1$. Thus according to (\ref{a13}) the field equations on $\Sigma_0$ are
\begin{eqnarray}\label{e4d}
3H^{2}&=&8\pi G\rho _{eff},\\
\label{ee4d}
2\frac{\ddot{a}}{a}+H^{2}&=&-8\pi G p_{eff},
\end{eqnarray}
where we have defined the effective energy density and pressure by $\rho _{eff}=\rho_{(IM)}+\Lambda _0$ and $p_{eff}=p_{(IM)}-\Lambda _0$.
In terms of the warping factor $A(\psi)$ the energy density $\rho _{(IM)}$ and the pressure $p_{(IM)}$ of induced matter  are determined by
\begin{equation}\label{b13}
\rho_{(IM)}=\overset{\star\star}{A}-2\overset{\star}{A}^{2},\qquad p_{(IM)}=-(\overset{\star\star}{A}-2\overset{\star}{A}^{2}) ,
\end{equation}
where both quantities must be evaluated at $\psi=\psi_0$. Using the equation (\ref{b7}) the previous $\rho_{(IM)}$ and $p_{(IM)}$ become
\begin{equation}\label{b14}
\rho_{(IM)}=\frac{n(1-2n)}{4\psi_0^2},\qquad p_{(IM)}=-\frac{n(1-2n)}{4\psi_0^2}.
\end{equation}
It can be easily seen from the first of these equations that in order to have a positive energy density $\rho_{(IM)}$ necessarily $0<n<1/2$. Taking into account this restriction and in order to warranty a positive value of $\varphi$ for every value of $\psi$ in the equation (\ref{b10}), $n=1/4$ seems to be a convenient election of $n$. In this case the Brans-Dicke scalar field given by the equation (\ref{b10}) and consequently the effective 5D gravitational coupling determined by (\ref{b11}) become respectively
\begin{eqnarray}\label{b15}
\varphi(\psi)&=&\frac{1}{16\pi G_{N}}\left(\frac{\psi +\psi_0}{2\psi_{0}}\right)^{2},\\
\label{b16}
G_{eff}(\psi)&=&16\pi G_{N}\left(\frac{2\psi_0}{\psi+\psi_0}\right)^{2}.
\end{eqnarray}
The effective equation of state parameter $\omega _{eff}$ in this case gives
\begin{equation}\label{b17}
\omega _{eff}\equiv\frac{p_{eff}}{\rho_{eff}}=\frac{p_{(IM)}-\Lambda_{0}}{\rho_{(IM)}+\Lambda_{0}}=-1,
\end{equation}
which corresponds to an equation of state for a purely cosmological constant. Thus we have on $\Sigma_0$ the effective cosmological constant
\begin{equation}\label{add1}
\Lambda_{eff}=\frac{n(5+6n)}{4\psi_0^2}
\end{equation}
 Finally this effective cosmological constant according to (\ref{b12}) for $n=1/4$ is given by
\begin{equation}\label{b18}
\Lambda_{eff}=\frac{13}{32\psi_{0}^{2}}.
\end{equation}
The present value of the cosmological constant according to cosmological estimations is approximately $\Lambda _{0}=10^{-54}\,m^{-2}$ or $\Lambda=10^{-47}\,GeV^{4}$ \cite{Rebi26}, thus when we impose this restriction to (\ref{b18}) it can be easily shown that $\psi _{0}=6.373\cdot 10^{26}\, m$. This value is slightly larger than  the present Hubble radius $\lambda _{H_0}\simeq 1.27\cdot 10^{26}\,m$, which give approximately the radius of our observable universe, so this could explain why for  observers located on $\Sigma_0$ the extended extra coordinate remains unobserved, at least directly.

\section{A dynamical cosmological constant $\Lambda (t)$CDM}

In order to obtain a dynamical cosmological constant in the present formalism, we use the 5D line element
\begin{equation}\label{dcc1}
dS_{5}^2 =\left(\frac{\psi}{\psi_0}\right)^{2\alpha}\left[dt^2-a^2(t)\delta_{ij}dx^{i}dx^j\right]-\Phi^2(t)d\psi^2,
\end{equation}
where $a(t)$ is the cosmic scale factor and $\Phi(t)$ is a metric function. Therefore, the field equations (\ref{ab1}) now read 
\begin{small}
\begin{eqnarray}
&& \psi^{2}\Phi^2(H^2-H\frac{\dot{\Phi}}{\Phi})-\alpha (2\alpha-1)\left(\frac{\psi}{\psi_0}\right)^{2\alpha}\nonumber\\ 
\label{dcc2}
&& =\frac{\psi}{6}\left(\frac{\psi}{\psi_0}\right)^{2\alpha}\left[\omega \psi\left(\frac{\overset{\star}{\varphi}}{\varphi}\right)^2-\alpha\left(\frac{\overset{\star}{\varphi}}{\varphi}\right)\right],\\
\label{dcc3}
&& \overset{\star}{\varphi}+\frac{3\alpha}{\psi}\varphi=0,\\
&& \psi^2\Phi^2\left(-2\frac{\ddot{a}}{a}-H^2-2H\frac{\dot{\Phi}}{\Phi}-\frac{\ddot{\Phi}}{\Phi}\right) +3\alpha(2\alpha-1)\left(\frac{\psi}{\psi_0}\right)^{2\alpha}\nonumber \\
\label{dcc4}
&&=-\frac{\psi}{2}\left(\frac{\psi}{\psi_0}\right)^{2\alpha}\left[\omega\psi\left(\frac{\overset{\star}{\varphi}}{\varphi}\right)^2-\alpha\frac{\overset{\star}{\varphi}}{\varphi}\right],\\
&& 2\alpha^2\left(\frac{\psi}{\psi_0}\right)^{2\alpha}-\psi^2\Phi^2\left(\frac{\ddot{a}}{a}+H^2\right)\nonumber\\
\label{dcc5}
&& =\psi^2\left(\frac{\psi}{\psi_0}\right)^{2\alpha}\left[\frac{\omega}{2}\left(\frac{\overset{\star}{\varphi}}{\varphi}\right)^2+\frac{\overset{\star\star}{\varphi}}{\varphi}\right].
\end{eqnarray}
\end{small}
The equation (\ref{a4}) evaluated on the metric  (\ref{dcc1}) can be written as
\begin{equation}\label{dcc6}
\overset{\star\star}{\varphi}+\frac{4\alpha}{\psi}\overset{\star}{\varphi}=0.
\end{equation}
Now, the combination of equations 2$\times$(\ref{dcc2})-(\ref{dcc4})+2$\times$(\ref{dcc5}) leaves to the equation
\begin{eqnarray}
&& \Phi^2\left(H^2+\frac{\ddot{\Phi}}{\Phi}\right)-\frac{\alpha(6\alpha-5)}{\psi^2}\left(\frac{\psi}{\psi_0}\right)^{2\alpha}\nonumber\\
\label{dcc6}
&& =\left(\frac{\psi}{\psi_0}\right)^{2\alpha}\left[\frac{11\omega}{6}\left(\frac{\overset{\star}{\varphi}}{\varphi}\right)^{2}-\frac{5\alpha}{6\psi}\frac{\overset{\star}{\varphi}}{\varphi}+2\frac{\overset{\star\star}{\varphi}}{\varphi}\right].
\end{eqnarray}
Separating variables in the equation (\ref{dcc6}) it results
\begin{eqnarray}\label{dcc7}
\left(\frac{\ddot{\Phi}}{\Phi}+H^2\right)\Phi^2=\gamma^2,\\
\left(\frac{\psi}{\psi_0}\right)^{2\alpha}\left[2\frac{\overset{\star\star}{\varphi}}{\varphi}+\frac{11\omega}{6}\left(\frac{\overset{\star}{\varphi}}{\varphi}\right)^2-\frac{5\alpha}{6\psi}\left(\frac{\overset{\star}{\varphi}}{\varphi}\right)\right]\nonumber\\
\label{dcc8}
+\frac{\alpha(6\alpha-5)}{\psi^2}\left(\frac{\psi}{\psi_0}\right)^{2\alpha}=\gamma^2,
\end{eqnarray} 
with $\gamma$ being a separation constant with $[length]^{-1}$ units. For a power law expanding universe with $H(t)=p/t$ a particular solution for (\ref{dcc7}) is
\begin{equation}\label{dcc9}
\Phi(t)=\Phi_0\left(\frac{t}{t_0}\right),
\end{equation}
where $t_0$ denotes the present time, $\Phi_0=\Phi(t_0)$ is a dimensionless constant and the relation $p=\gamma t_0/\Phi_0$ holds. We can normalize this solution by choosing $\Phi_0=1$, which is the value of the metric function $\Phi$ in the present time $t_0$.\\

The equations (\ref{dcc3}) and (\ref{dcc6}) have the common solution
\begin{equation}\label{dcc10}
\varphi (\psi)=M_p^2\left(\frac{\psi_0}{\psi}\right)^3,
\end{equation}
where $M_p$ is the Planckian mass. Inserting (\ref{dcc10}) in (\ref{dcc8}) we obtain the condition
\begin{equation}\label{dcc11}
\frac{11}{2}\frac{5+3\omega}{\psi_0^2}=\gamma^2.
\end{equation}
In a power law expanding universe $p=H_0t_0$, thus it is valid the relation $\gamma=\Phi_0 H_0$. Using the expressions (\ref{dcc9}) and (\ref{dcc10}) the dynamical cosmological constant reads
\begin{equation}\label{dcc12}
\Lambda(t)=\Lambda_0\left(\frac{t_0}{t}\right)^2,
\end{equation}
where
\begin{equation}\label{dcc13}
\Lambda_0=\frac{9\omega}{2\psi_0^2}.
\end{equation}
On the other hand, the 4D dynamical field equations induced on the hypersurface $\Sigma_0$ can be written as
\begin{eqnarray}\label{dcc14}
3H^2&=& 8\pi G \rho_{(IM)}+\Lambda(t)\\
\label{dcc15}
2\frac{\ddot{a}}{a}+H^2&=& -8\pi Gp_{(IM)}+\Lambda(t),
\end{eqnarray}
where $\rho_{(IM)}=T^{(IM)}\,^t_t$ and  $p_{(IM)}=-(1/3)(T^{(IM)}\,^x_x+T^{(IM)}\,^y_y+T^{(IM)}\,^z_z)$ with the help of (\ref{a12}) read
\begin{eqnarray}\label{imden}
\rho_{(IM)}&=&2\left[\left(\frac{\dot{\Phi}}{\Phi}\right)^2+\frac{\ddot{\Phi}}{\Phi}\right]-\frac{3}{\psi_0^2 \Phi^4},\\
\label{impre}
p_{(IM)}&=&-\left(2H\frac{\dot{\Phi}}{\Phi}-\frac{3}{\psi_0^2\Phi^4}\right),
\end{eqnarray}
 which for (\ref{dcc9}) acquire the form
\begin{eqnarray}\label{dcc16}
\rho_{(IM)}&=& \frac{2}{t^2}-\frac{3t_0^4}{\psi_0^2\Phi_0^4}\frac{1}{t^4},\\
\label{dcc17}
p_{(IM)}&=& -\frac{2p}{t^2}+\frac{3t_0^4}{\psi_0^2\Phi_0^4}\frac{1}{t^4},
\end{eqnarray}
where in order to have a energy density  $\rho_{(IM)}>0$, the condition $\psi_0\gg (\sqrt{3}p_0^2/\Phi_0^2) r_{H_0}^2$ must be satisfied. Thus, it follows from (\ref{dcc14}), (\ref{dcc15}), (\ref{dcc16}) and (\ref{dcc17}) that the total equation of state parameter (EOS) can be written as
\begin{equation}\label{dcc18}
\omega_T=\frac{p_{T}}{\rho_T}=\frac{\omega_{IM}\Omega_{(IM)}-\Omega_{\Lambda}}{\Omega_{(IM)}+\Omega_{\Lambda}},
\end{equation}
where $\Omega_{(IM)}=\rho_{(IM)}/p_{(IM)}$ is the density parameter of induced matter, $\Omega_{\Lambda}=\Lambda(t)/(8\pi G\rho_{c})$ and $\omega_{IM}$ is the EOS parameter for induced matter, being $\rho_c$ the critical ener\-gy density. Now, as the induced matter accounts for the matter sources in the universe, it is natural to consider the separation 
\begin{equation}\label{dcc19}
\Omega_{(IM)}=\Omega_{m}+\Omega_{r},
\end{equation}
with $\Omega_{m}$ describing the density parameter for matter and $\Omega_r$ is the density parameter for radiation. Therefore, the EOS parameter for induced matter can be written in the form
\begin{equation}\label{dcc20}
\omega_{IM}=\frac{\omega_m\Omega_m+\omega_r\Omega_r}{\Omega_m+\Omega_r},
\end{equation}
where we have regarded barotropic EOS for matter and radiation. It is not difficult to show that for matter obey a dust EOS parameter i.e. $\omega_m =0$, the equation (\ref{dcc17}) implies that the power of scale factor of the universe is determined by
\begin{equation}\label{dcc21}
p=\frac{3(H_0t_0)^2}{2(\psi_0H_0)^2\Phi_0^4}-\frac{1}{6H_0^2}\rho_{c_0}\Omega_{r_0}(H_0t_0)^2,
\end{equation}
where the subindex $0$ denotes evaluated in the present time. It is a straightforward calculation to see that when the condition
\begin{equation}\label{dcc22}
\frac{3(H_0 t_0)^2}{2H_0^2\psi_0^2}\left(\frac{1}{1+\frac{1}{6H_0^2}(H_0 t_0)^2\rho_{c_0}\Omega_{r_0}}\right)>\Phi_0^2,
\end{equation}
is satisfied, we have $p>1$, which means that on this limit the expansion of the universe will be accelerated.\\

The deceleration parameter reads
\begin{equation}\label{dcc23}
q=\frac{1}{2}\left[1-3\frac{\left(2p+\frac{\Lambda_0 (H_0t_0)^2}{8\pi G H_0^2}\right)-\frac{3(H_0t_0)^4}{\phi_0^2(H_0\Phi_0)^4}\frac{1}{t^2}}{\left(2+\frac{\Lambda_0(H_0t_0)^2}{8\pi GH_0^2}\right)-\frac{3(H_0t_0)^4}{\psi_0^2(H_0\Phi_0)^4}\frac{1}{t^2}}\right].
\end{equation}
Now, it follows from (\ref{dcc12}) and (\ref{dcc13}) that as $\psi_0=6.373\cdot 10^{26}m$, in order to $\Lambda_0$ satisfy the observed value today $\Lambda\simeq 10^{-54}m^{-2}$, it requires that the BD parameter has the value $\omega=0.09025$. \\

Thus, with the help of the equations (\ref{dcc12}), (\ref{dcc13}), (\ref{dcc16}), (\ref{dcc17}), (\ref{dcc18}), (\ref{dcc19}) and (\ref{dcc20}), the total equation of state parameter can be written in the form
\begin{equation}\label{dcc24}
\omega_T=\frac{-\frac{2p}{t^2}+\frac{3t_0^4}{\psi_0^2\Phi_0^4}\frac{1}{t^4}-\frac{\Lambda_0}{8\pi G}\left(\frac{t_0}{t}\right)^2}{\frac{3}{t^2}-\frac{3t_0^4}{\psi_0^2\Phi_0^4}\frac{1}{t^4}+\frac{\Lambda_0}{8\pi G}\left(\frac{t_0}{t}\right)^2},
\end{equation}
which for the present time reduces to
\begin{equation}\label{dcc25}
\omega_{T_0}=\frac{-\frac{2pH_0^2}{(H_0t_0)^2}+\frac{3}{\psi_0^2\Phi_0^4}-\frac{\Lambda_0}{8\pi G}}{\frac{2H_0^2}{(H_0t_0)^2}-\frac{3}{\psi_0^2\Phi_0^4}+\frac{\Lambda_0}{8\pi G}}.
\end{equation}
The expression (\ref{dcc25}) can be used to obtain the value of $\Phi_0$ in terms of $\omega_{T_0}$, which results to be
\begin{eqnarray}\label{dcc26}
\Phi_0&=&\pm\gamma\left(\frac{H_0t_0}{\psi_0}\right)^{1/2},\\
\label{dcc27}
\Phi_0&=&\pm i\gamma\left(\frac{H_0t_0}{\psi_0}\right)^{1/2},
\end{eqnarray}
where 
\begin{equation}\label{dcc28}
\gamma=\left[\frac{24\pi G (1+\omega_{T_0})}{(1+\omega_{T_0})(H_0t_0)^2\Lambda_0+16\pi G(p+\omega_{T_0})H_0^2}\right]^{1/4}.
\end{equation}
Of course we are interested in the physical real solutions (\ref{dcc26}). According to the data combination Planck+WP+BAO+SN, in the present time the EOS parameter ranges in the interval $\omega_0=-1.10^{+0.08}_{-0.07}$ \cite{Obs1}. These values are achieved by (\ref{dcc25}) when in agree with (\ref{dcc26}) $\Phi_0$ have the values 
\begin{equation}\label{dcc29}
\Phi_0=\pm 0.01054883615^{+1.06\cdot 10^{-8}}_{-5.017\cdot 10^{-8}}.
\end{equation}
Finally, taking into account (\ref{dcc29}), it follows from (\ref{dcc23}) that the present decceleration parameter is bounded by
\begin{equation}\label{dcc30}
q_0=-1.15^{+0.012}_{-0.105}.
\end{equation}
In this manner we have shown that if we consider that gravity propagates along the 5D bulk obeying a Brans-Dicke theory, then a dynamical cosmological constant can be induced on our 4D universe in concordance with obser\-vational evidences.

\section{Final Comments}

In this letter ragarding our 4D observable universe as  locally and isometrically  embeded
into a 5D spacetime, where gravity is governed by a Brans-Dicke (BD) theory in vacuum, we have obtained cosmological solutions in which the cosmological constant, that can vary with time, arises in the 4D field equations as an induced geometrical term. In this sense in this model the acce\-le\-rating expansion of the universe is explained without the need to introduce dark energy.\\

The main condition that makes the embedding possible is that the BD scalar field must be depending solely of the extra coordinate and $\varphi(\psi_0)=(16\pi G)^{-1}$.  In here our 4D observable universe is represented by a generic hypersurface $\Sigma:\psi=\psi_0$, member of a foliation of the 5D ambient space. The field equations induced on the 4D space-time are the usual Einstein field equations plus an extra term that can be written as a dynamical cosmological constant term.\\

 We found that there exist solutions to the 5D field equations that induce a $\Lambda CDM$ model in 4D (when $g_{\psi\psi}$ is time independent) or a $\Lambda(t)$ CDM model (when $g_{\psi\psi}$ has time dependence). Matter sources in 4D are geometrically induced from the 5D geometry (see the energy-momentum tensor of induced matter (\ref{a12})). Thus we obtain in 4D a modified theory of general relativity.\\
 
Finally, in order to illustrate how the model works, we have obtained two cosmological solutions. One that describes a 4D universe with an induced cosmological constant and other one in which the cosmological cons\-tant varies with time. The first one of these scenarios results of considering for the 5D space-time the class of geometries known as warped product geometries. In this example, 4D sources of matter are also geome\-trically induced as it is usually done in the Wesson's  induced matter approach. The value of the cosmological constant geometrically induced agrees with its value determined by observations $\Lambda _{0}=10^{-54}\,m^{-2}$  when $\psi _{0}=6.373\cdot 10^{26}\, m$. The cosmological constant in this example is effective in the sense that the induced matter EOS parameter also corresponds to a cosmological constant.\\

In the second example, we have derived a more general metric solution of the 5D field equations that induce on the 4D space-time a dynamical cosmological constant of the form  $\Lambda(t)=\Lambda_0(t_0/t)^{-2}$. We have shown that this solution describes a universe with accelerating expansion when $\Phi_0=\pm 0.01054883615^{+1.06\cdot 10^{-8}}_{-5.017\cdot 10^{-8}}$. Moreover, with this solution a $\Lambda(t)CDM$ model can be reproduced. In the both examples the fifth coordinate evaluated on $\Sigma _0$ must be $\psi_0\simeq 5.018\cdot \lambda_{H_0}$, where $\lambda _{H_0}\simeq 1.27\cdot 10^{26}\,m$ is the present Hubble radius.  The fact that $\psi_0$ is slightly larger than the Hubble radius, could explain why in this  model, for hypothetical observers located on $\Sigma _0$  the extended fifth extra dimension remains unobserved, at least directly.\\

\section*{Acknowledgements}

\noindent  J.E.M.A and J. A. Licea acknowledge CONACYT M\'exico and Departamento de Matem\'aticas of Universidad de Guadalajara, for financial support. J. Zamarripa acknowledge Conacyt and Centro Universitario de los Valles for financial support.  A. Peraza acknowledge Departamento de F\'isica of  Universidad de Guadalajara, for financial support.

\end{document}